# Re-analysis of the Cassini RPWS/LP data in Titan's ionosphere. Part II: statistics on 57 flybys.

**A. Chatain[1,2], J.-E. Wahlund[3], O. Shebanits[3,4], L. Z. Hadid[2,3], M. Morooka[3], N. J. T. Edberg[3], O. Guaitella[2], and N. Carrasco[1]**

[1] Université Paris-Saclay, UVSQ, CNRS, LATMOS, Guyancourt, France.

[2] LPP, Ecole polytechnique, Sorbonne Université, Institut Polytechnique de Paris, CNRS, Palaiseau, France.

[3] Swedish Institute of Space Physics, Uppsala, Sweden.

[4] Imperial College London, United-Kingdom.

Corresponding author: Audrey Chatain (audrey.chatain@ens-paris-saclay.fr)

**Key Points:**

- Four cold electron populations are detected by the Cassini Langmuir probe in Titan's ionosphere below 1200 km altitude.

- Three populations are likely produced by particle precipitation ionization and photo-ionization in the atmosphere and on the probe boom.

- The fourth electron population, observed on dayside below 1200 km, could be photo- or thermo-emitted from dust grains.

**Abstract**

The ionosphere of Titan hosts a complex ion chemistry leading to the formation of organic dust below 1200 km. Current models cannot fully explain the observed electron temperature in this dusty environment. To achieve new insight, we have re-analyzed the data taken in the ionosphere of Titan by the Cassini Langmuir probe (LP), part of the Radio and Plasma Wave Science package. A first paper (Chatain et al., n.d.) introduces the new analysis method and discusses the identification of 4 electron populations produced by different ionization mechanisms. In this second paper, we present a statistical study of the whole LP dataset below 1200 km which gives clues on the origin of the 4 populations. One small population is attributed to photo- or secondary electrons emitted from the surface of the probe boom. A second population is systematically observed, at a constant density (~500 cm$^{-3}$), and is attributed to background thermalized electrons from the ionization process of precipitating particles fom the surrounding magnetosphere. The two last populations increase in density with pressure, solar illumination and EUV flux. The third population is observed with varying densities at all altitudes and solar zenith angles except on the far nightside (SZA > ~140°), with a maximum density of 2700 cm$^{-3}$. It is therefore certainly related to the photo-ionization of the atmospheric molecules. Finally, a fourth population detected only on the dayside and below 1200 km reaching up to 2000 cm$^{-3}$ could be photo- or thermo-emitted from dust grains.

# 1 Introduction

Titan's ionosphere is a complex environment, partly governed by an active ion chemistry leading to the formation of organic dust grains (Lavvas et al., 2013; Vuitton et al., 2019; Waite et al., 2007). From 2004 to 2017, the ionospheric plasma has been investigated *in situ* by many instruments on-board the Cassini mission, such as the ion and neutral mass spectrometer INMS (Cui et al., 2009; Magee et al., 2009; Mandt et al., 2012; Waite et al., 2005, 2007), the Cassini plasma spectrometer CAPS (Coates et al., 2007, 2009; Crary et al., 2009; Wellbrock et al., 2013, 2019) and the radio and plasma wave science (RPWS) package.

In particular, RPWS was designed to continue and improve the first radio and plasma wave measurements in the Saturn system, done by Voyager 1 and 2 (Gurnett et al., 1981; Scarf et al., 1982). This instrument aimed to study radio emissions, plasma waves, thermal plasma and dust (Gurnett et al., 2004). It was formed by a suite of antennas and sensors including a Langmuir probe. The Langmuir probe measures low energy charged particles in ionized environments, and took *in situ* measurements in the magnetosphere and ionosphere of Titan at the occasion of 126 close flybys. Electron density and temperature were deduced from these measurements (Ågren et al., 2009; Edberg et al., 2010, 2013b, 2015; Wahlund et al., 2005). However, electron temperature measurements are not well reproduced by models of the ionosphere of Titan below 1200 km: model results give electrons too cold by a factor of 2-3 (Galand et al., 2014; Mukundan & Bhardwaj, 2018; Richard et al., 2011; Shebanits et al., 2017a; Vigren et al., 2013, 2016).

In the objective to get additional clues of processes that are missing in the models, we re-analyzed the Langmuir probe dataset taken in the ionosphere. This dataset was sampled during the 13 years of the mission. In particular, we searched for clues about the production of electrons hotter than the model predictions. This work is presented in two parts. In a first paper (Chatain et al., n.d.), referred to as 'paper I', we detailed the method used for the re-analysis of the data, and in a second paper (this one), we present the results obtained for the complete Cassini dataset. Paper I showed that several electron populations are systematically measured by the Langmuir probe in the ionosphere of Titan. Depending on the solar illumination and the altitude probed, 2 to 4 populations with different densities, temperatures and potentials were identified.

The detection of several electron populations is not unusual. Previous works observed 2 to 3 electron populations in the upper atmosphere of Venus (Intriligator et al., 1979), in Mars ionosphere (Hanson & Mantas, 1988; Mitchell et al., 2000), in the ionized environment of the comet 67P/Churyumov-Gerasimenko (Eriksson et al., 2017), in Saturn's magnetosphere (Schippers et al., 2008) and at Enceladus (Tokar et al., 2006). In all these cases, a cold electron population, thermalized through collisions, cohabits with hot suprathermal electrons coming from the solar wind, the magnetosphere of Saturn or photo-ionization. The main difference in the case of the ionosphere of Titan is that all the four electron populations observed are cold (<1 eV). Indeed, in these conditions models predict that all electrons should be thermalized (Galand et al., 2014). Previous works studying Langmuir probe data in Titan's ionosphere (Ågren et al., 2009; Edberg et al., 2010, 2013b, 2015; Wahlund et al., 2005) already used 2 and sometimes 3 electron populations to fit the data, without investigating on their variations and origins.

To closely investigate the origins of the 4 populations detected, we analyzed the whole Langmuir probe dataset in the ionosphere of Titan and searched for correlations with the EUV flux (computed by Shebanits et al., 2017b) and the positive and negative ion densities (deduced from a multi-instrument study by Shebanits et al., 2016). Section 2 presents the Langmuir probe dataset, section 3 shows the electron density and temperature results, section 4 studies correlations and sections 5 discusses the origin of the four populations based on the results of the previous sections.

## 2 Materials and Methods

### 2.1 Langmuir probe data in Titan's ionosphere

The Langmuir probe (LP) was part of the Radio and Plasma Wave Science (RPWS) package on-board Cassini (Gurnett et al., 2004). In this study we used data from the voltage sweep mode, one of the three operational modes of the probe. In this mode, the voltage of the probe is swept between +4 and -4 V and the current collected is recorded. From this we can in particular deduce the electron density and temperature. Further details are given in paper I.

In this work, we focused on the region below 1200 km in altitude. We analyzed the 57 Cassini flybys that went below this altitude (T5, 16-21, 23, 25-30, 32, 36, 39-43, 46-51, 55-59, 61, 65, 70-71, 83-88, 91-92, 95, 100, 104, 106-108, 113, 117-121, 126). During these flybys, Cassini spent ~15 min below 1400 km, during which it continuously acquired voltage sweeps (26-30 in total, at different altitudes). The spatial resolution is limited to ~3 km due to the spacecraft motion at ~6 km/s. The flybys happened at various seasons, solar zenith angles (SZA) and latitudes. Their characteristics are represented in Figure 1. More precisely, the Cassini mission lasted 13 years and observed two seasons: it arrived in the system of Saturn in 2004 just after the northern hemisphere winter and stayed until the summer solstice in 2017. The vernal equinox marked the middle of the total mission in 2010.

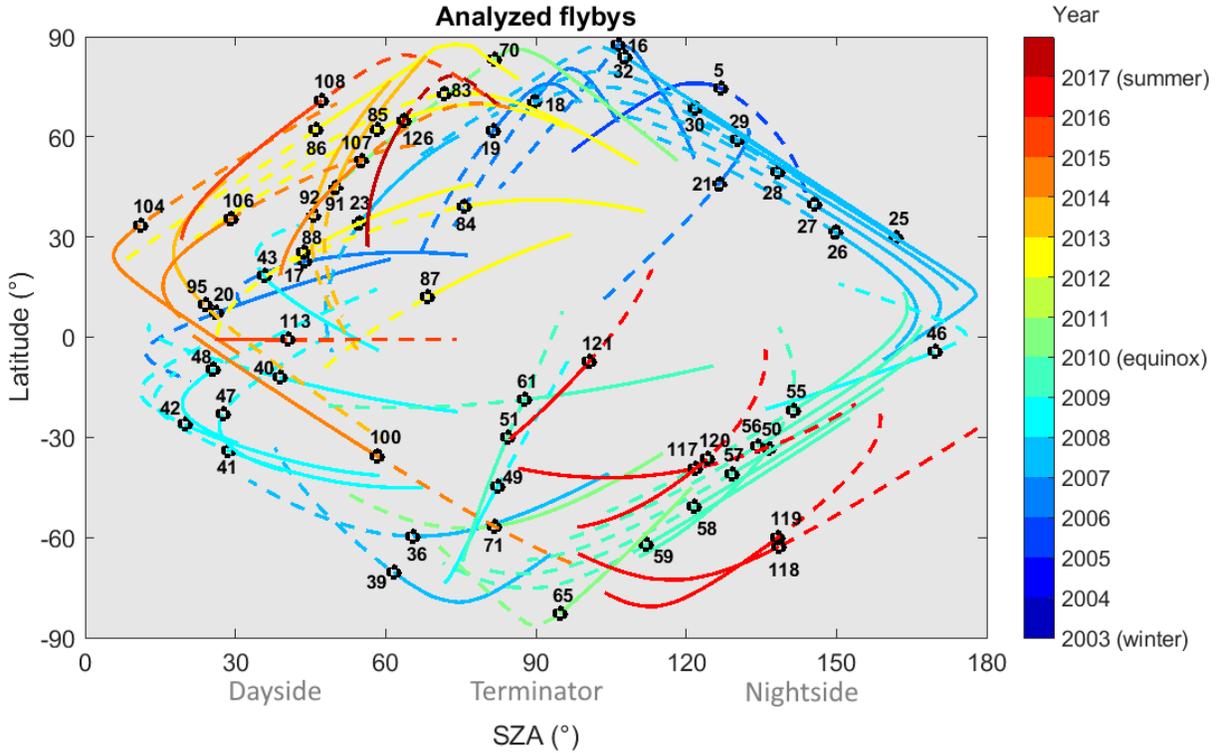

**Figure 1.** Trajectories of the 57 flybys analyzed as a function of solar zenith angle, latitude and year. Closest approach is marked by black circles. Inbound parts of flybys are given with plain lines and outbound parts with dashed lines. Seasons indicated in the colorbar correspond to the northern hemisphere.

## 2.2 Method to deduce electron density and temperature from voltage sweeps

The electron characteristics in the ionosphere are derived from the voltage sweep analysis. The current collected by the probe is due to positive ions (that give a negative current), negative ions, ambient electrons and photoelectrons emitted by the nearby surfaces (that give a positive current). We first remove the current due to ions with a linear fit. Then, the electron current is fitted assuming that each electron population has a Maxwellian speed distribution and using the Orbital Motion Limited (OML) theory with a Sheath Limited (SL) correction. The method is discussed in detail in paper I. Depending on the altitude and the solar illumination, 2 to 4 electron populations are detected, and their densities and temperatures are retrieved with generally 10% to 30% uncertainties (confidence interval at 95%). The lower detection limit of the probe in temperature is 0.015 eV (~175 K), due to the electrical work function for the probe coating material (TiN, ~15 mV rms) (Veszelei, 1997). The four detected populations are named $P_1$, $P_2$, $P_3$ and $P_4$ according to their increasing potentials from $P_1$ to $P_4$ (see paper I). The origins of these populations are discussed at the end of the paper (Section 6). Similarly to previous works (Ågren et al., 2009; Gustafsson & Wahlund, 2010; Wahlund et al., 2009; Wang et al., 2015), we attributed the sampled $P_1$ population to photoelectrons emitted from the probe boom.

## 3 Results: electron densities and temperatures

### 3.1 Statistics on 57 flybys

Figure 2 gives the results on electron density and temperature obtained for each of the 4 populations for the 57 flybys studied. Each of the four populations have a different behavior with altitude, which are described below.

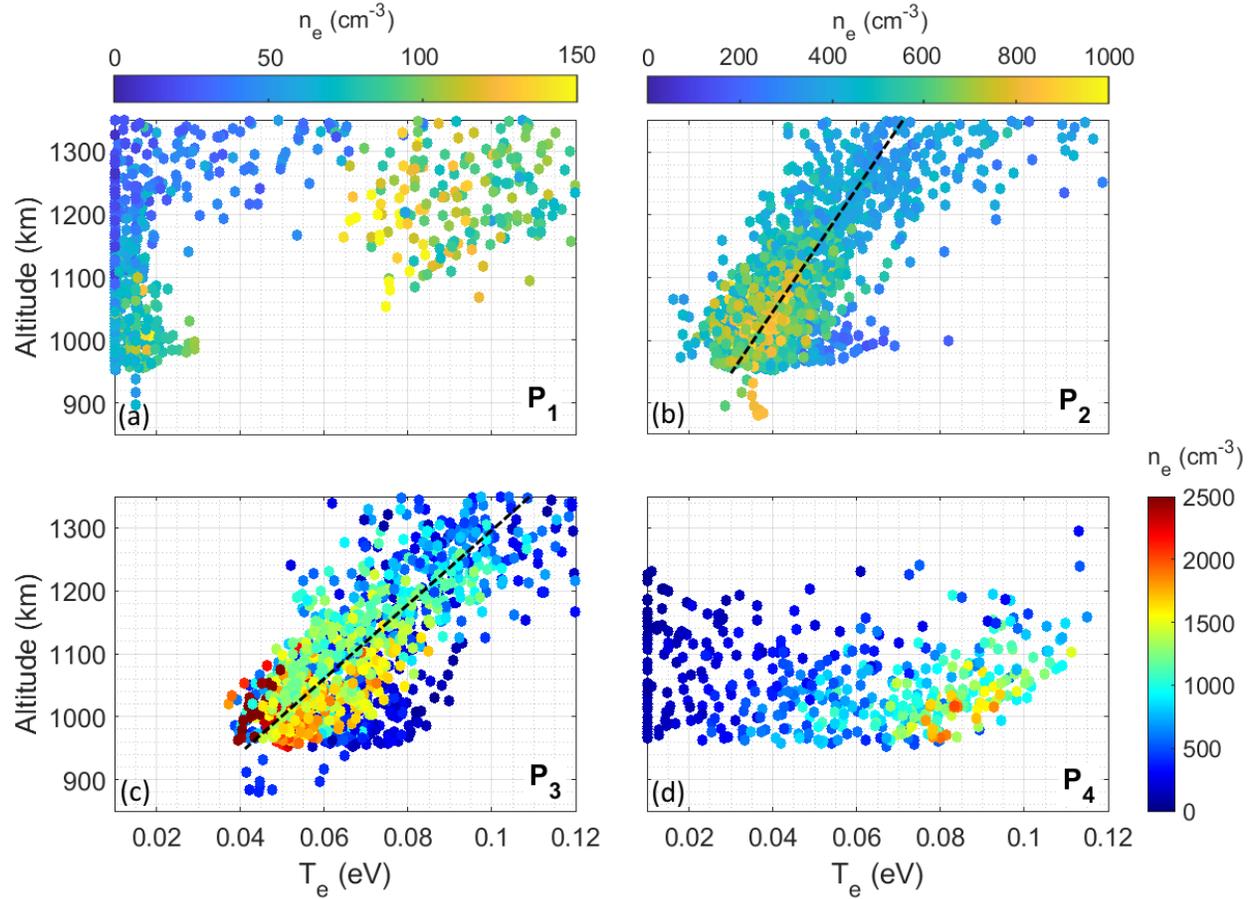

**Figure 2.** Electron temperature and density as function of altitude for the four electron populations: (a) $P_1$, (b) $P_2$, (c) $P_3$ and (d) $P_4$. Data from 57 flybys. Linear altitude trends are indicated with a black dashed line in the cases of $P_2$ and $P_3$ (respectively at -0.010 eV/100 km and -0.017 eV/100 km).

$P_1$ (Figure 2a) is a population with a very low density ($< 150$ cm$^{-3}$). Two different behaviors are observed: the low temperature electrons ($< 0.02$ eV, 230 K) are found at all altitudes, and hotter electrons ($0.07 - 0.12$ eV; 810-1390 K) are found only above 1100 km. These could be due to two different formation processes (electrons emitted by the probe boom after collision with a photon or with an energetic particle). However, $P_1$ electron properties are deduced from a very small current, whose structure is hard to analyze in more detail. In particular, it is difficult to conclude on the existence of formation processes that could occur simultaneously.

$P_2$ electrons (Figure 2b) reach temperatures mainly between 0.025 and 0.08 eV (290-930 K), with a maximum at 0.12 eV (1390 K). $P_3$ electrons (Figure 2c) are generally hotter, with a minimum at 0.04 eV (460 K). $P_2$ and $P_3$ show in average a linear decrease in temperature with a decreasing altitude. The altitude trends for the two populations are different: estimations give -0.010 eV (115 K) / 100 km for $P_2$ and -0.017 eV (200 K) / 100 km for $P_3$. Therefore, the electrons of $P_3$ are hotter, but their temperature decreases with altitude ~70% faster than the electrons of $P_2$. Higher densities are globally found at lower altitudes.

$P_4$ electrons (Figure 2d) have the particularity to appear only below ~1250-1200 km altitude, with a large range of temperatures, between 0.01 and 0.12 eV (120-1390 K). Higher densities are correlated with higher temperatures.

### 3.2 Variations of the electron density with SZA and altitude

#### 3.2.1 Trends for the four populations

The electron populations have a strong dependence on Solar Zenith Angle (SZA). Figure 3 shows the results in terms of electron density.

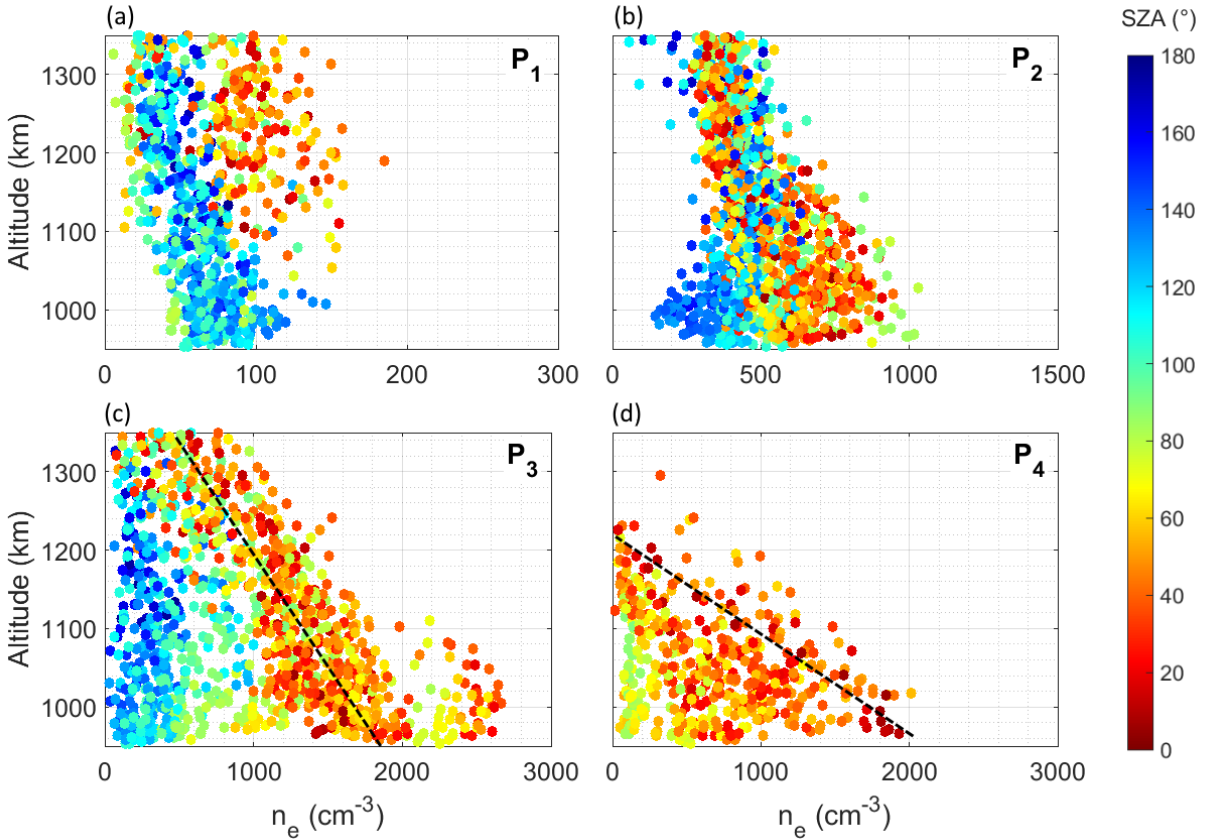

**Figure 3.** Electron density as a function of altitude and Solar Zenith Angle (SZA) for the four electron populations: (a) $P_1$, (b) $P_2$, (c) $P_3$ and (d) $P_4$. Data from 57 flybys. Linear altitude trends are indicated with a black dashed line in the cases of $P_3$ and $P_4$ (respectively at +350 cm$^{-3}$/100 km and +770 cm$^{-3}$/100 km).

$P_1$ electrons (Figure 3a) at low SZA are only observed at high altitudes. Their density remains low (~100 cm$^{-3}$) in any condition. $P_2$ electron density (Figure 3b) stays globally constant around 500 cm$^{-3}$ at all altitudes and all SZA. Only a small shift is observed at low altitude, with higher densities on the dayside (~700 cm$^{-3}$) and lower densities on the nightside (~400 cm$^{-3}$). Later we will show that the lack of dependence of population 2 on SZA is an indication of a source that is not related to solar illumination, like the thermalization of magnetospheric electrons.

The effect of SZA is stronger on $P_3$ and $P_4$. While the density of $P_3$ (Figure 3c) is globally constant with altitude on the nightside (~300 cm$^{-3}$), it strongly increases with decreasing altitude on the day side, with a slope of +350 cm$^{-3}$ / -100 km (up to ~1900 cm$^{-3}$ at 950 km). For $P_4$ (Figure 3d), the maximum density reached at each altitude increases strongly with decreasing altitude and SZA. In the case of SZA < 20°, the increase of density is about +770 cm$^{-3}$ / -100 km. The increase of the total electron density at low solar zenith angle was already observed by Ågren et al. (2009) on flybys from T16 to T42. The re-analysis of the complete dataset confirmed their observation and showed that the total density increase is due to the increase of $P_3$ electrons and the apparition of $P_4$ electrons. This indicates that $P_3$ and $P_4$ have an origin linked to solar illumination. Nevertheless, their different behaviors with altitude prove that their formation mechanisms are distinct. In particular, we will see later that the presence of negative ions/dust at lower altitude could be related to the formation of $P_4$ electrons at lower altitude.

### 3.2.2 Repartition of the negative charge carriers

The four electron populations have densities varying with altitude and SZA. Figure 4 synthesizes the average repartition of negative charge carrier densities (electrons and negative ions/dust grains) with altitude and SZA. Negative ion/dust grain values are from Shebanits et al. (2016). In Titan's ionosphere, negative ions can reach very large masses and finally become dust grains of few nanometers (Lavvas et al., 2013; Waite et al., 2007). As dust grains are also negatively charged in this environment, we usually do not make a distinction between large negative ions and dust grains. Shebanits et al. (2016) deduced the negative ion/dust grain density by combining the mass/charge distributions measured by the Cassini spectrometers CAPS/IBS, CAPS/ELS and INMS, and the Langmuir probe measurement of the ion current. Above 1050 km in altitude, the negative charges essentially come from the electrons. On the opposite, below 1000 km, the negative ions/dust grains play a dominant role. Besides, as their average charge is estimated to be >1 (Shebanits et al., 2016), the negative charge born by the negative ions/dust grains could be higher compared to electrons for a same density. In the repartition of the electrons, a strong difference appears between the dayside and the nightside. $P_2$ electrons dominate the nightside whereas $P_3$ electrons play the main role on dayside.

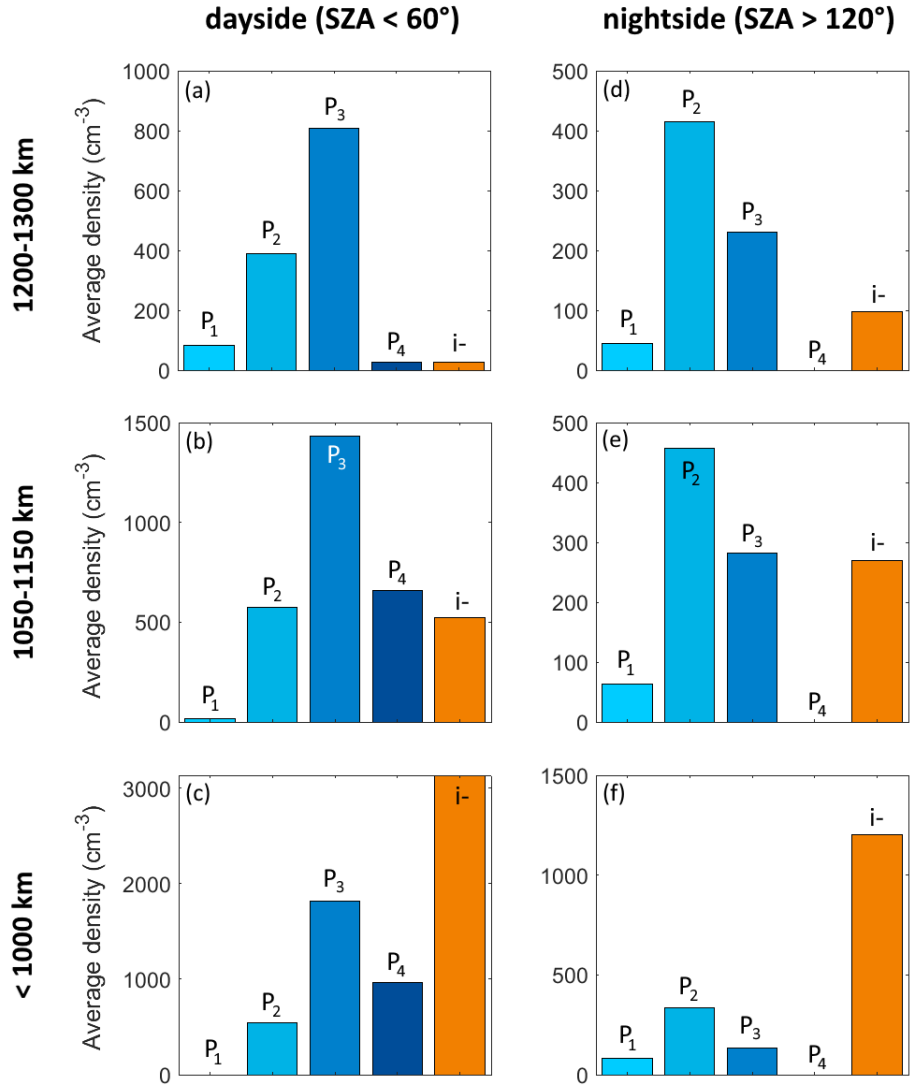

**Figure 4.** Average densities of negative charge carriers: the four electron populations ($P_1$, $P_2$, $P_3$, $P_4$) and the negative ions/dust grains (i-). Comparison at different altitudes, (a,b,c) on dayside (SZA < 60°) and (d,e,f) on nightside (SZA > 120°).

### 3.3 Variations of the electron temperature with SZA and altitude

Figure 5 shows the variation of electron temperature with SZA for each of the four populations. At higher altitudes, above 1250 km, no strong differences are observed with SZA. However, with decreasing altitudes from 1250 to 850 km, a differentiation occurs between dayside and nightside measurements.

Concerning population $P_1$ (Figure 5a), there is a neat separation between two blocks below 1050 and 1250 km, that we name C1 and C2. C1 corresponds to cold electrons (< 0.03 eV, 350 K) present on nightside, down to the lower altitudes reached by the spacecraft. On the dayside, C2 hotter electrons (0.07-0.12 eV, 810-1390 K) are observed, but only above 1050 km. However, this can be due to the fact that the collected current created by $P_2$ and $P_3$ electrons increase strongly and possibly cover the current due to $P_1$ electrons, which can thus not be detected (see paper I).

As is evident from Figure 2, the electrons of populations $P_2$ (Figure 5b) and $P_3$ (Figure 5c) get colder with decreasing altitude. However, at a given altitude, the temperatures of $P_2$ and $P_3$ are globally constant with SZA. This was previously observed with the analyses of one single electron population (Ågren et al., 2009). Only a small difference is observed below 1050 km: electrons on nightside seems slightly hotter (+0.01-0.02 eV), for SZA > 130° in the case of $P_2$, and for SZA > 100° in the case of $P_3$. Nevertheless, these values are within the fit error bars.

In the case of $P_4$, Figure 5d shows a sharp limit at 90° SZA: no electrons from the $P_4$ family can be observed on the nightside. Therefore, $P_4$ electrons appear only in sunlit regions. However, there is no relation between the electron temperature and the SZA.

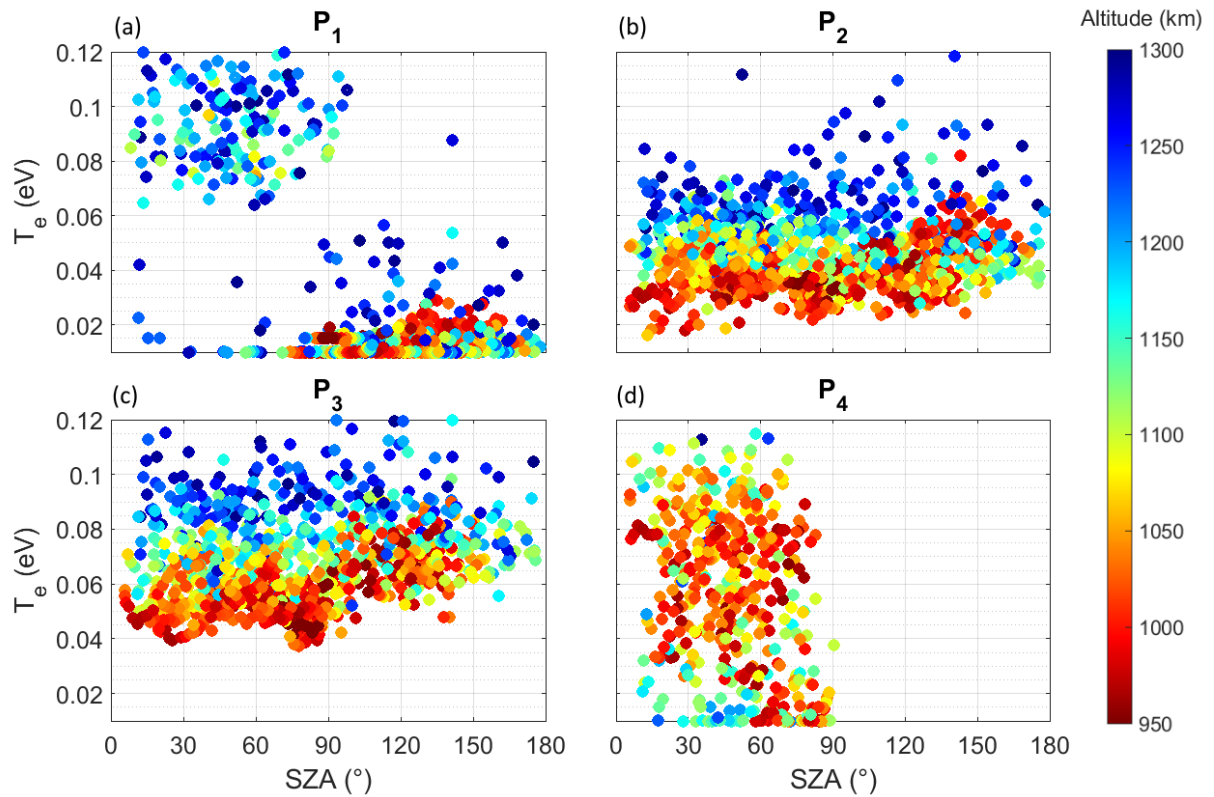

**Figure 5.** Electron temperature as a function of SZA and altitude for the four electron populations: (a) $P_1$, (b) $P_2$, (c) $P_3$ and (d) $P_4$. Data from 57 flybys.

### 3.4 Relation between the $P_4$ electron density and temperature

At a given altitude, electrons from $P_2$ and $P_3$ do not show any variation in temperature with density. For those populations, the temperature seems essentially governed by altitude. However, results are very different for population $P_4$. Figure 6 shows a linear trend between electron temperature and density, observed at all altitudes where $P_4$ electrons are detected. The linear coefficient varies with altitude from +0.016 eV (190 K) / 100 cm$^{-3}$ at 1150-1250 km (Figure 6a) to +0.006 eV (80

K) / 100 cm⁻³ below 1050 km (Figure 6c) for SZA < 70°. Below 1050 km (Figure 6c), the slope varies also with SZA, and near the terminator (SZA > 70°) the coefficient is higher (+0.011 eV (130 K) / 100 cm⁻³).

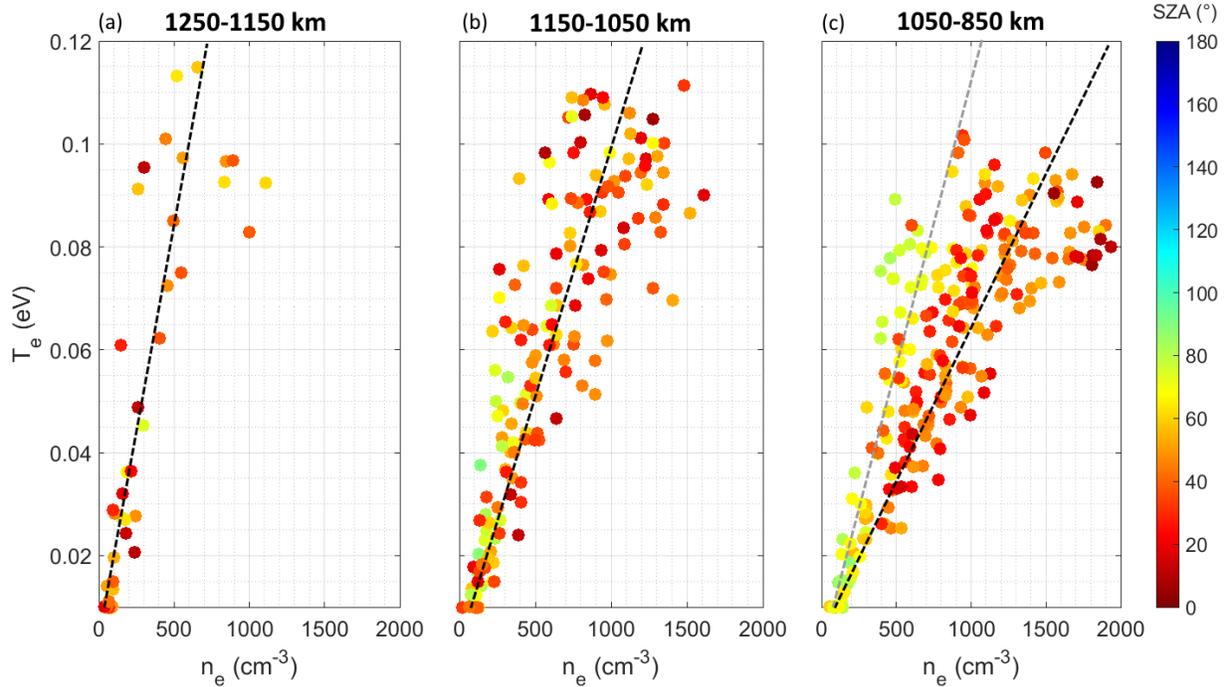

**Figure 6.** Electron temperature as a function of density and Solar Zenith Angle (SZA) in the case of population P4 at 3 different altitudes: (a) 1250-1150 km, (b) 1150-1050 km and (c) 1050-850 km. Data from 34 flybys on dayside. Linear trends are indicated with a black dashed line for SZA < 70° [(a) +0.016 eV/100 cm⁻³, (b) +0.010 eV/100 cm⁻³ and (c) +0.006 eV/100 cm⁻³], and with a grey dashed line for SZA > 70° below 1050 km [(c) +0.011 eV/100 cm⁻³].

# 4 Correlations with EUV fluxes, seasons and ion densities

## 4.1 Correlations with EUV fluxes

Extreme ultraviolet (EUV) fluxes used in this work are computed by Shebanits et al. (2017b) at the distance of Saturn from the solar irradiance measurements by the Solar EUV Experiment (SEE) instrument from the Thermosphere Ionosphere Mesosphere Energetics and Dynamics (TIMED) mission and the Solar Stellar Irradiance Comparison Experiment (SOLSTICE) from the Solar Radiation and Climate Experiment (SORCE) mission (https://lasp.colorado.edu/lisird/). Therefore, EUV measurements in this study refer to the unattenuated solar EUV flux at the top of Titan's ionosphere.

Correlations observed with EUV fluxes depends on the population chosen. P₂ is globally not EUV dependent. This is consistent with the origin of P₂ suggested above, as the thermalization of magnetospheric electrons. P₁ electron temperature on dayside slightly increases with the EUV flux (+0.01 eV (115 K) every 0.1x10⁻⁴ W.m⁻²), as shown on Figure 7a. This is also consistent with their formation by photoemission on the probe boom.

A varying EUV flux has also a strong effect on $P_3$ (plotted below 1000 km on Figure 7b) and $P_4$ (Figure 7c) densities, confirming their dependence on solar irradiation. Below 1000 km, $P_3$ density increases roughly linearly, of +300 $cm^{-3}$ every $0.1x10^{-4}$ $W.m^{-2}$. In parallel, $P_4$ denser cases are only observed at higher EUV fluxes.

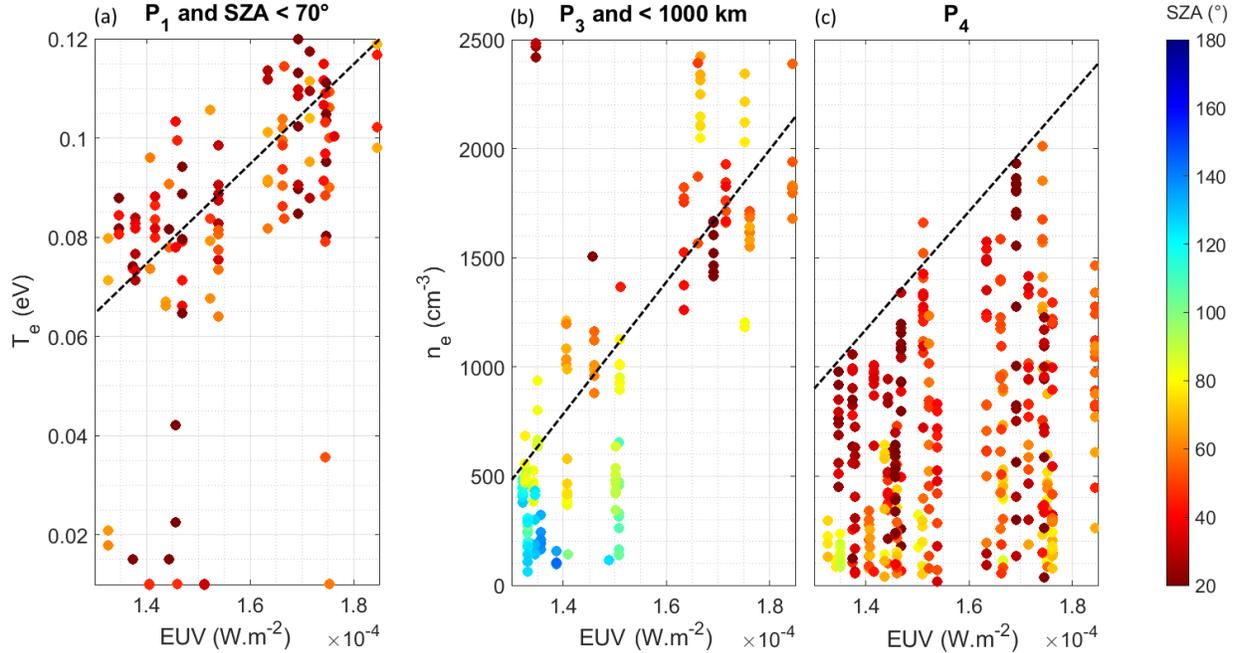

**Figure 7.** Electron characteristics as a function of extreme UV flux (integrated between 0.5-160.5 nm). (a) $P_1$ electron temperature for SZA < 70°. (b) $P_3$ electron density below 1000 km. (c) $P_4$ electron density. Linear trends are indicated with a black dashed line: (a) +0.01 $eV/0.1x10^{-4}$ $W.m^{-2}$, (b) +300 $cm^{-3}/0.1x10^{-4}$ $W.m^{-2}$ and (c) +275 $cm^{-3}/0.1x10^{-4}$ $W.m^{-2}$.

### 4.2 Impact of seasons and solar cycles

The Cassini mission covered two seasons in the Saturnian system, from the beginning of winter in the northern hemisphere to the summer solstice (May 2017). The Langmuir probe took measurements all over the mission, which enables to study the evolution of electron density and temperature with seasons (Edberg et al., 2013b; Shebanits et al., 2017b).

The previous sections show that SZA affect electron populations, changing their densities and temperatures. Therefore, it is necessary to compare results at different seasons obtained with similar SZA. Figure 8a presents SZA during flybys as a function of seasons (latitude and date). In the case of SZA > 70°, two blocks of data have sufficient data points to be compared: one from 2006 to 2008 in the southern hemisphere (further named 'B1') and one from 2012 to 2014 in the northern hemisphere ('B2').

Figure 8b shows the evolution of the $P_1$ electron temperature for SZA < 70°. A higher temperature is observed in the 'B2' part than in the 'B1' part. Neither $P_2$ density nor temperature at a given altitude shows variations with latitude and seasons. Concerning $P_3$ electrons, their temperature is not affected by seasons and latitude, but their density in the cases where SZA < 70° is globally higher in the 'B2' part than in 'B1'. Figure 8d illustrates this effect with data points acquired below 1100 km. As for the population $P_4$, a similar effect is seen on both the electron density (Figure 8f) and temperature, as a linear correlation links its density and temperature (see Section 3.4).

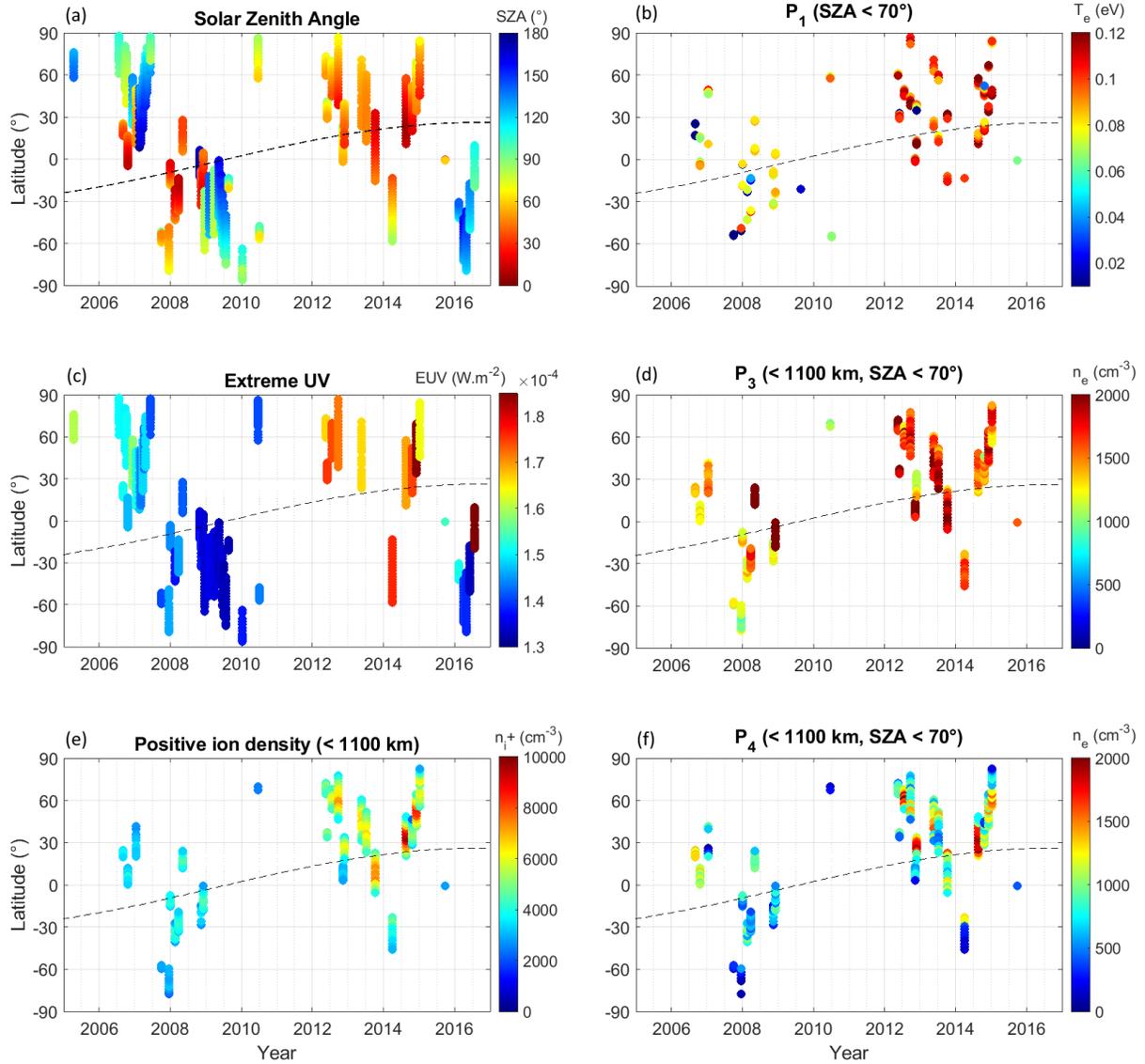

**Figure 8.** Evolution with seasons: (a) solar zenith angle during the flybys, (b) $P_1$ electron temperature on dayside, (d,f) $P_3$ and $P_4$ electron densities on dayside below 1100 km, (c) EUV flux (integrated between 0.5-160.5 nm) and (e) positive ion density below 1100 km (data from Shebanits et al., 2017b). The sub-solar latitude is indicated in black dashed line (in the northern hemisphere, seasons change from winter, to spring equinox in 2010, to summer).

In conclusion, on dayside, $P_3$ and $P_4$ electrons are denser in the northern hemisphere from 2012 to 2014 (northern spring) than in the southern hemisphere from 2006 to 2008 (southern summer). Similarly, $P_1$ and $P_4$ electrons are hotter during the northern spring than during the southern summer. These observations are anti-correlated with the evolution of the Saturn-Sun distance, which was at its closest in 2003 (9 A.U.) and at its furthest in 2018 (10 A.U.).

Figure 8c presents the EUV flux and shows a good correlation with Figure 8b and 8d. The observed effect is then certainly linked to the variations of the EUV flux with years. In addition, these observations are correlated with the solar cycle that was at a minimum between 2005 and 2011 and at a maximum between 2011 and 2016. In conclusion, the modifications of the $P_1$ and $P_4$ electron temperatures and $P_3$ and $P_4$ densities in long time scales is mainly due to the evolution of EUV flux along the solar cycle, which has a greater influence than seasons. This effect of the solar cycle on the total electron density has already been observed in 2013 (Edberg et al., 2013b).

Finally, Figure 8e shows the evolution of positive ion density measured by Shebanits et al. (2017b) plotted similar to the graphs for $P_3$ and $P_4$ electrons in Figure 8d and 8f for easy comparison. We note similarities with $P_3$ and $P_4$, suggesting a correlation between these electron populations, the ion density and the EUV flux. Besides, a strong dependence between ion densities and the EUV flux (and then the solar cycle) has been observed by Shebanits et al. (2017b).

### 4.3 Correlations with ion density

The positive and negative ion/dust grain densities have previously been retrieved by Shebanits et al. (2016, 2017b), from the comparative study of different instruments on-board Cassini: the RWPS/LP, the Ion and Neutral Mass Spectrometer INMS, the Cassini Plasma Science Electron Spectrometer CAPS/ELS and Ions Beam Spectrometer CAPS/IBS. We use these results in the current section.

$P_1$ and $P_2$ electron densities do not show any correlation with the positive ($n_i+$) and negative ($n_i-$) ion densities, contrary to $P_3$ and $P_4$. Figure 9 plots the $P_3$ and $P_4$ electron densities as a function of the positive ion density. The SZA (a,c) and altitude (b,d) are indicated to help the interpretation. The two last subfigures (e,f) show the negative ion/dust grain density.

Three different domains can globally be distinguished depending on the positive ion density.

(1) For $n_i+ < 2300$ cm$^{-3}$ (usually at high altitudes or on nightside), $P_3$ electron density is linearly correlated to the positive ion density with a linear coefficient $n_i+/n_e(P_3)$ close to 2: there are two positive ions for one $P_3$ electron. In these cases, there are no $P_4$ electrons.

(2) For a positive ion density between ~2300 and ~3500 cm$^{-3}$ (mainly on dayside and below 1150 km), the $P_3$ electron density reaches a plateau. The $P_4$ electrons appear and are linearly correlated to the positive ion density with a linear coefficient $n_i+/n_e(P_4)$ close to 2.

(3) For $n_i+ > 3500$ cm$^{-3}$ (mainly on dayside and below 1075 km), both $P_3$ and $P_4$ electron densities are globally constant with an increasing positive ion density. In these conditions, the positive ion charge density is no longer compensated by the electrons ($P_1$ and $P_2$ having low densities), but by negative ions that increase strongly. Indeed, Figure 9 e-f shows a linear coefficient $n_i+/n_i-$ close to 1. We observe that these cases are all at the lowest altitudes and at higher ion and dust densities.

Ions can bear several charges to the contrary of electrons. Nevertheless, from the above results we can still conclude that low positive ion densities on nightside or at high altitude on dayside are mainly compensated by $P_3$ electrons (from photo-ionization). Higher ion density cases, usually on dayside below 1150 km, are strongly related to $P_4$ electrons (likely emitted by dust grains). Finally, in the highest density cases, on dayside below 1075 km, the increase of positive ions at lower altitudes is mainly compensated by a simultaneous increase of negative ions.

An exception to these 3 domains is observed in a few flybys at lower altitudes (< 975 km) on nightside, where $P_3$ electrons are at low density (< 500 cm$^{-3}$), $P_4$ electrons are absent, but still positive and negative ions are present in high quantities. These cases certainly show remnants of ions formed on dayside (Shebanits et al., 2016).

Concerning electron temperatures, we observed no dependence with the ion densities. For $P_2$ and $P_3$, the temperature depends only on the altitude, and for $P_4$ the temperature depends on the $P_4$ density.

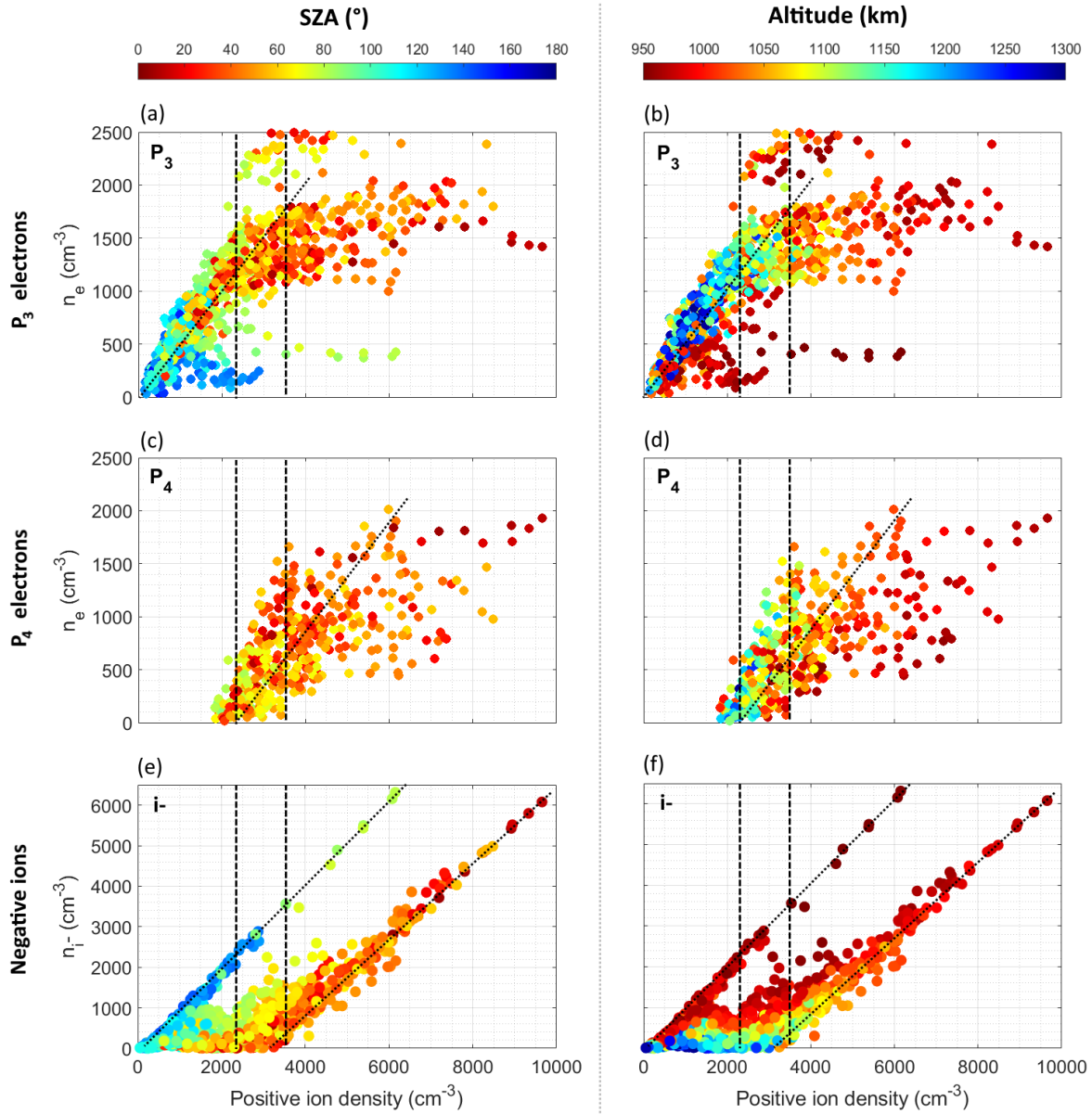

**Figure 9.** Densities of $P_3$ electrons (a,b), $P_4$ electrons (c,d) and negative ions/dust grains (e,f) as a function of positive ion density. Colors refer to (a,c,e) Solar Zenith Angle and (b,d,f) altitude. Positive and negative ion densities are from Shebanits et al. (2016, 2017b). Initial linear trends are indicated in black dotted lines. Trend transition zones are plotted in black dashed lines (2300 and 3500 cm$^{-3}$).

## 5 Investigation of the higher electron density cases

This section focuses on the ionospheric peak for each flyby, defined as the altitude where the total electron density measured is at its maximum. For a few flybys the ionospheric peak is maybe not reached by Cassini. In these cases, we take the lower altitude where Cassini went for the plots. An example of such a case is T104, and its density profile can be seen in Figure 11 of paper I.

Figure 10 shows the maximum of the total electron density as a function of Solar Zenith Angle (SZA) for the 57 flybys analyzed. The altitude of the maximum density detected is indicated with the colorbar. This graph can be compared to the Figure 6 in Ågren et al. (2009), and to the Figure 3a in Edberg et al. (2013a), here completed with a larger dataset. We observe the same trends: densities are low and constant on the nightside and increase at lower SZA. On the dayside the ionospheric peak is at lower altitude. Edberg et al. (2013a) observed that T85 presents unsually high densities. We do the same observation. Some other flybys show also high densities at the ionospheric peak. We selected T91, T104 and T107 in this study.

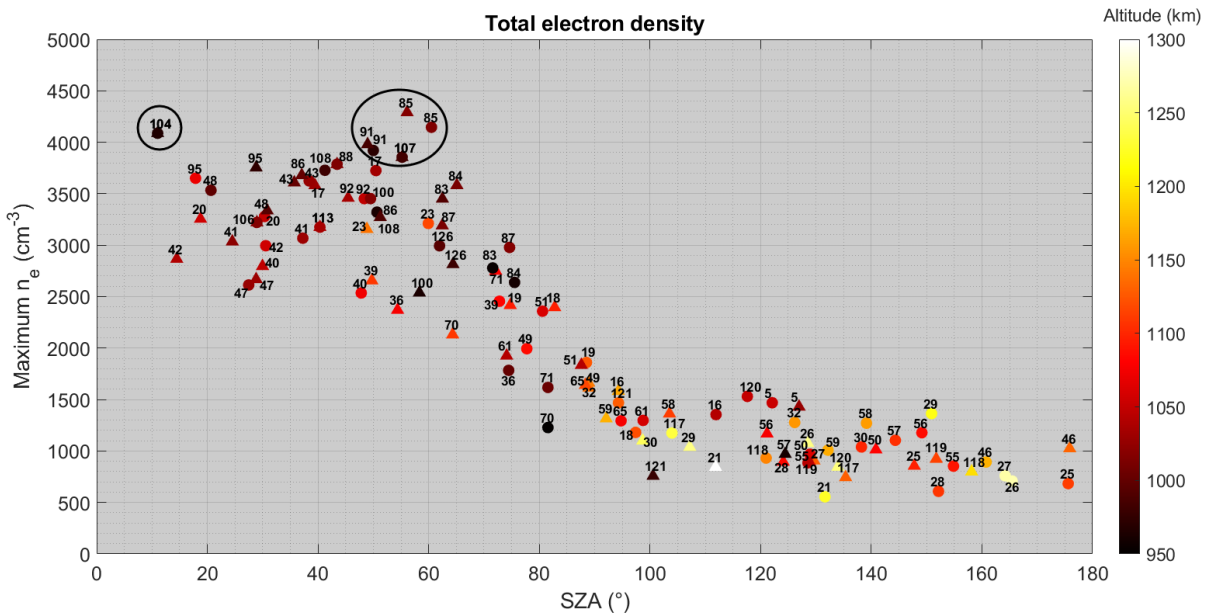

**Figure 10.** Total electron density at the ionospheric peak as a function of Solar Zenith Angle (SZA) for the 57 flybys going below 1300 km. The numbers of the flybys are indicated on the plot. Dots correspond to the inbound part of a flyby, and triangles to outbound. The altitude of the ionospheric peak is indicated in the colorbar. 4 flybys at unusually high densities are circled in black.

To have an insight on the origin of these high density cases, we plotted clones of Figure 10, with colorbars indicating various parameters. No clear correlation is observed with the Saturn Local Time and the RAM angle (see Figure 11a-b). Therefore, these high density cases cannot directly be explained by the position of Titan in Saturn's magnetosphere or by the position of Cassini relatively to Titan's motion. Electron densities in the ionosphere below 1200 km are globally not influenced by the magnetospheric conditions in usual cases. A strong correlation is observed between the high density cases and the extreme UV flux (see Figure 11c). We conclude that the high electron density cases are mainly due to extreme UV fluxes. In the case of T85, it could be linked to a coronal mass ejection (CME) happening simultaneously (Edberg et al., 2013a). Such high density cases are also concomitant with high negative ion densities (see Figure 11d).

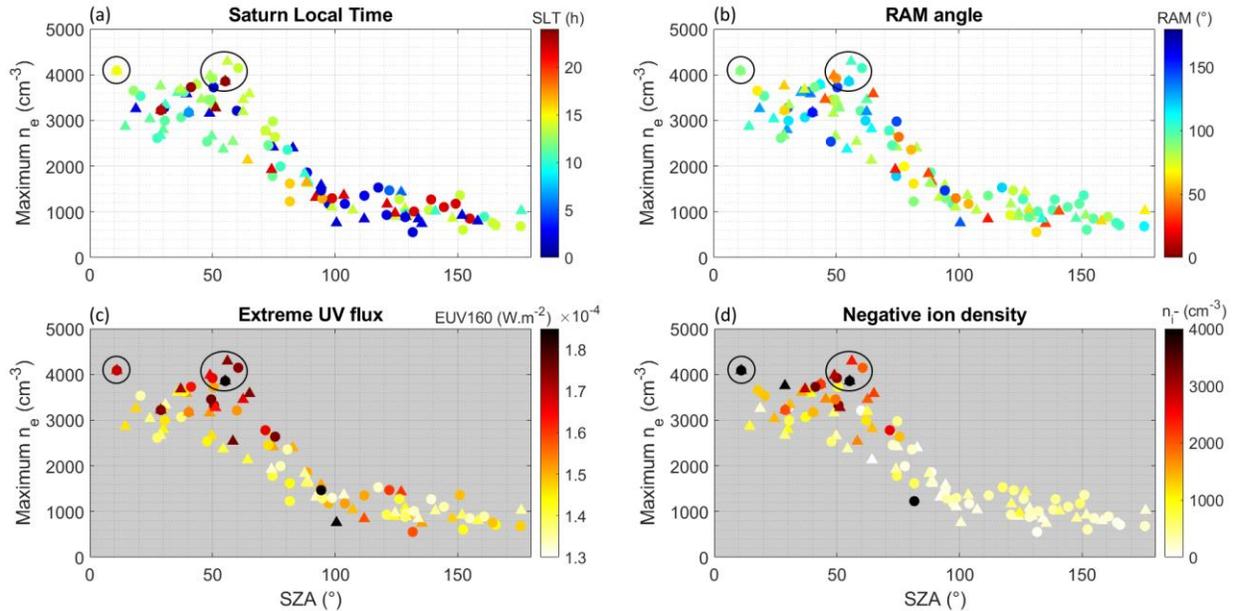

**Figure 11.** Total electron density at the ionospheric peak as a function of Solar Zenith Angle (SZA) for the 57 flybys going below 1300 km. Colorbars refer to different variables: (a) Saturn Local Time, (b) RAM angle, (c) extreme UV flux (from Shebanits et al. (2017b), see Section 4.1) and (d) negative ion density (from Shebanits et al. (2016, 2017b), see Section 4.3). Dots correspond to the inbound part of a flyby, and triangles to outbound. 4 flybys at unusually high densities are circled in black.

We investigated further the origin of high density cases by looking at the electron density results for the populations $P_2$, $P_3$ and $P_4$ at the ionospheric peak (see Figure 12 a-b-c). We observed that these cases do not show particularly high densities for $P_2$ and $P_3$. However, they are the cases with higher $P_4$ densities. Looking at density profiles (example for T85 in Figure 12d, profiles for the other flybys are very similar), we concluded that in these high density cases, the electron density increase at lower altitude is mainly due to $P_4$ electrons. These results are coherent with previous observations showing that $P_4$ electrons are strongly correlated with EUV fluxes and negative ion densities (see Section 4).

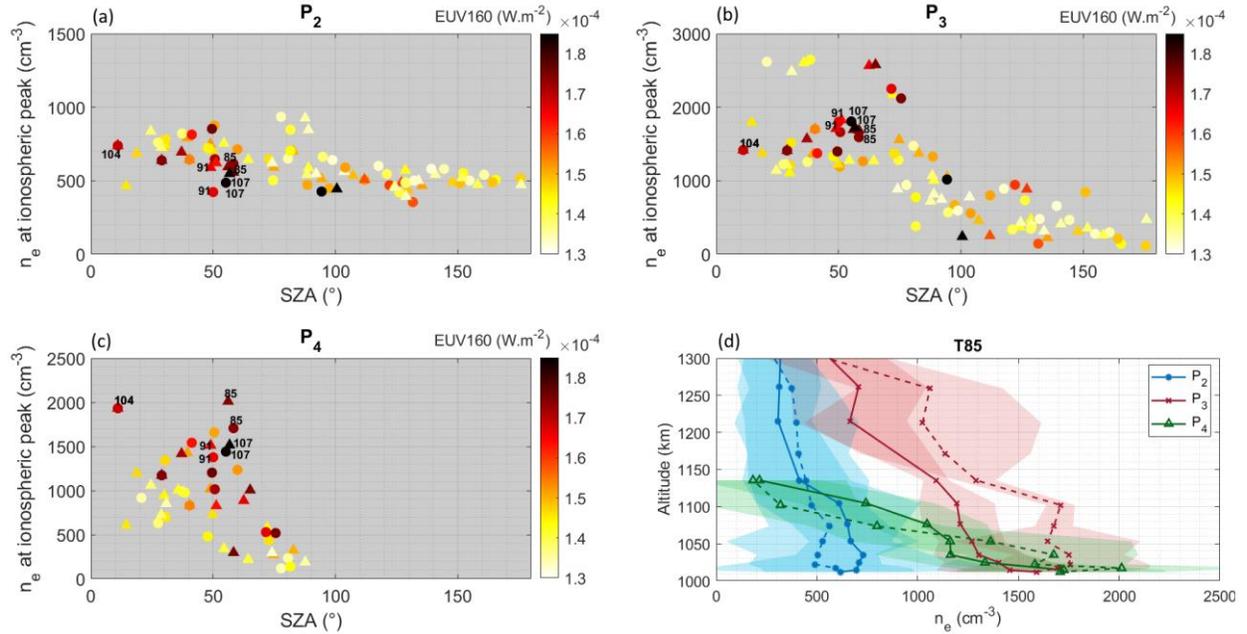

**Figure 12.** Electron density for population $P_2$ (a), $P_3$ (b) and $P_4$ (c) at the ionospheric peak as a function of Solar Zenith Angle (SZA) and extreme UV flux (from Shebanits et al. (2017b), see Section 4.1). Dots correspond to the inbound part of a flyby, and triangles to outbound. The 4 flybys at usually high densities in Figures 10 and 11 are indicated on the plots. (d) Electron density for $P_2$, $P_3$ and $P_4$ as a function of altitude measured during the T85 flyby. This subfigure is similar to Figure 11b in paper I, showing the electron density profiles in the case of T104. The fit confidence intervals at 95% are indicated in color shaded area (see paper I for more details). Inbound data points are linked with a plain line, and outbound data points with a dashed line.

# 6 Origins of the electron populations

## 6.1 Summary of the characteristics of the 4 populations

Table 1 summarizes the correlations observed in the previous sections between $n_e$, $T_e$, altitude, solar irradiation and ion density, for the 4 populations. Its content is described in Section 6.2, where it helps to suggest origins for the 4 populations.

**Table 1.** Summary of the correlations between $n_e$, $T_e$, altitude, solar irradiation and ion density, for the 4 populations.

| Popula-tion | $n_e$, $T_e$ | altitude | SZA, EUV, seasons | ions | suggested origin |
|---|---|---|---|---|---|
| $P_1$ | **$\underline{n_e}$: constant and low density** ($\sim$80 cm$^{-3}$) **$\underline{T_e}$: 2 cases:**, **C1** (<0.02 eV) **C2** ($\sim$0.1 eV) | **case C1**: $P_1$ electrons are present at all altitudes **case C2**: $P_1$ electrons are present only above 1100 km | **C1**: found only on nightside **C2**: always present (and only) on dayside, $T_e$ is hotter at higher EUV fluxes. | / | electrons are emitted from the probe boom by bombardment of photons (for **C2**) or energetic particles (for **C1**) |
| $P_2$ | / | **$\underline{n_e}$: globally constant density** (500 cm$^{-3}$), with a slight shift at lower altitudes depending on the SZA (400-700 cm$^{-3}$) **$\underline{T_e}$: linear decrease of the temperature with decreasing altitude** (-0.01 eV / -100 km) | **$\underline{n_e}$**: at low altitude, $P_2$ electrons are slightly denser on dayside (x2). | / | thermaliza-tion of the suprathermal electrons formed by particle precipitation |
| $P_3$ | / | **$\underline{n_e}$**: constant on nightside. **Density linearly increases on dayside with decreasing altitude** (+350 cm$^{-3}$ / -100 km) **$\underline{T_e}$: linear decrease with decreasing altitude** (-0.017 eV / -100 km) | **$\underline{n_e}$: denser on dayside (x6+).** Below 1000 km, $P_3$ electrons are denser at higher EUV fluxes. | **$\underline{n_e}$: linear increase** of $n_e$ with increasing $n_i^+$ ($n_i+/n_e = 2$) for $n_i+ < 2300$ cm$^{-3}$; constant $n_e$ for larger $n_i^+$. | from photo-ionization |
| $P_4$ | $n_e$ and $T_e$ proportional ($\sim$0.01 eV / 100 cm$^{-3}$) | **Exists only below 1200-1150 km.** Denser cases are observed at lower altitudes. | **Always present (and only) on dayside.** Denser cases are only at high EUV fluxes. | **$\underline{n_e}$: linear increase with $n_i^+$** ($n_i+/n_e = 2$) for $n_i+ < 4000$ cm$^{-3}$; constant for larger $n_i^+$. | from interaction of solar photons with aerosols or heavy negative ions |

6.2 Suggested origins for the four populations

We discuss here possible origins for the four populations. They are schematized in Figure 13.

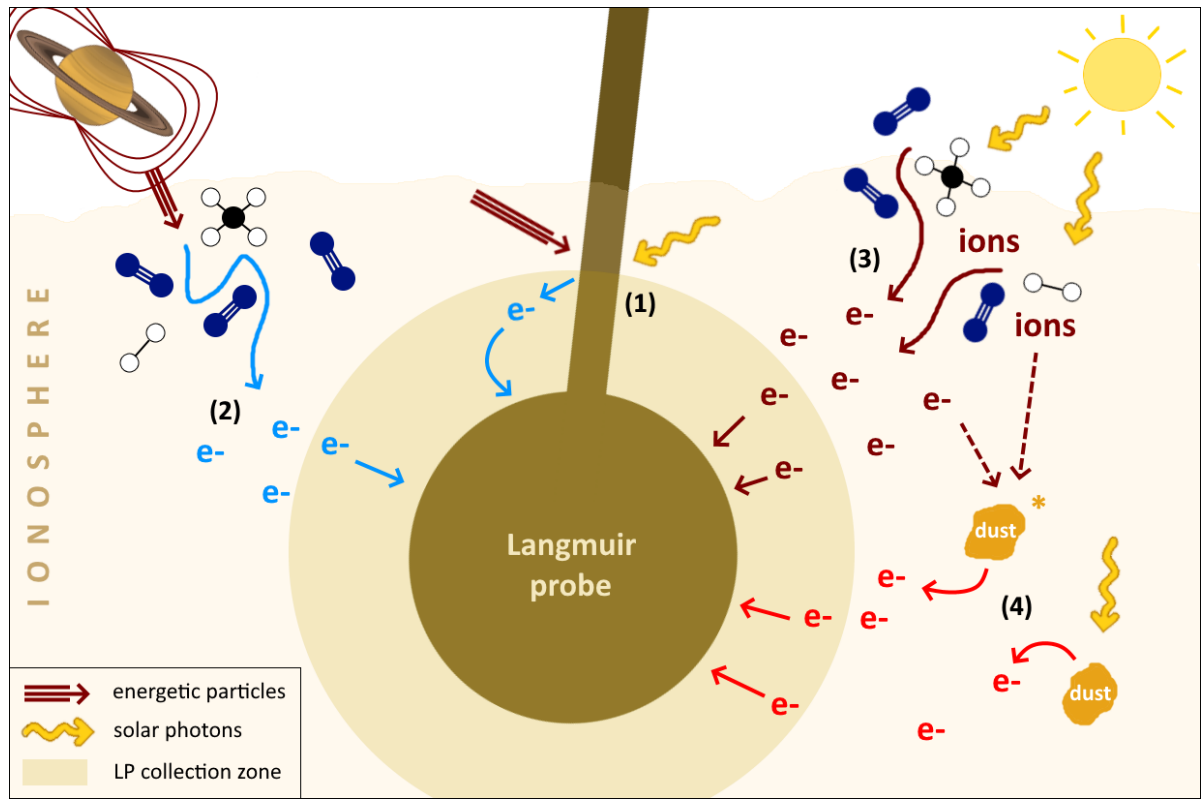

**Figure 13.** Scheme representing suggested origins for the 4 electron populations detected by the Langmuir probe in Titan's ionosphere. * next to a dust grain stands for 'excited'.

$P_1$ electrons are nearly always present and have a very low density. Their floating potential being the smallest of all the populations (see paper I), it is a strong indication that they are formed on the probe boom, emitted after collision with energetic photons or particles. Their higher temperature on dayside suggests that photo-emission forms more energetic electrons at the surface of the boom than secondary electrons emitted after collision with an energetic particle. On nightside, $P_1$ electron temperature is close to the 0.015 eV instrumental limitation, which could also affect the detection of this population.

The $P_2$ electron population depends only slightly on solar illumination. It has a constant density (~500 cm$^{-3}$) at all altitudes and solar zenith angles. It is always present, and is the major ion population on nightside. Its temperature decreases linearly with decreasing altitude, down to ~0.04 eV (460 K) at 1000 km. For these reasons, it is likely that $P_2$ electrons are background thermalized electrons, present in all the ionosphere of Titan. Their constancy with solar illumination indicates that they could mainly be due to the thermalization of suprathermal magnetospheric electrons initially formed by particle precipitation.

$P_3$ and $P_4$ populations show strong correlations with the SZA and the EUV flux. Consequently, their origin must be linked to the solar irradiation. In addition, both have a density increasing with decreasing altitude. To summarize, increases in solar photons and atmospheric pressure (i.e. neutral density) lead to an enhanced formation of $P_3$ and $P_4$ electrons: these electrons are formed from the energy deposit of solar photons in the atmosphere. The main difference between the two populations is their temperature. $P_3$ electron temperature depends essentially on the altitude: a higher collision rate at higher pressure lead to a lower temperature. In contrast, the temperature of $P_4$ electrons depends mainly on their density: their temperature is higher when their density is higher. These points suggest a different origin for the two populations. $P_3$ electrons nearly appear at all altitudes, whereas $P_4$ electrons are confined below 1150-1200 km, where negative ions and dust particles are dominant. At the light of these observations, we suggest that $P_3$ electrons come from the photo-ionization of neutral atmospheric molecules (mainly $CH_4$ and $N_2$), forming suprathermal photoelectrons and colder secondaries, while $P_4$ electrons are formed on the dust grains or heavy negative ions by interaction with the solar photons.

### 6.3 Discussion on the dust origin of $P_4$

$P_4$ electrons are a clue of the interaction of the plasma with the aerosols in the ionosphere of Titan. Extracted from the aerosols, their floating potential is different from the plasma potential. This explains why the $P_4$ electrons are collected at a very different voltage bias of the probe than the other populations (see paper I).

A few hypotheses can be made on the processes forming electrons from the interaction of aerosols grains with solar photons. Photo-extraction from the bulk of the material requires strong UV photons (> 6 eV), while photo-desorption can happen with low energy photons (1-2 eV), but with lower efficiency. In these cases, the energy of the electron formed depends on the energy of the incident photon. This can explain the linear dependence between $n_e$ and $T_e$: higher and stronger solar radiations lead to a higher quantity of $P_4$ electrons, and at higher energies. Tigrine et al. (2018) studied photo-emission of electrons on analogues of Titan's aerosols under VUV irradiation. Using an extrapolation of their data points, we observe that the formation of electrons at 0.1 eV would be possible and require incoming photons of ~6 eV.

Woodard et al. (2020) studied a laboratory dusty plasma with a Langmuir probe and also observed an unexpected electron population at a potential different from the plasma potential. They showed that the potential of the unexpected population is the same as the potential of the aerosols. Using numerical modeling they investigated its origin. In their conditions, the electron photo-emission and the secondary electron emission from the flux of charged species to the particle surface are not sufficient to explain the density of the new population. They propose another process: the thermo-emission of electrons from the nanoparticles, enhanced by electrostatic effects. Indeed, the aerosols are heated by the ion bombardment, the collection of electrons and the recombination of radicals and ions at their surface. The accumulated heat cannot be transferred to the background gas because of the low pressure and the inefficient radiation of small particles. On the other hand, the charge of the aerosols decreases the barrier for electron emission. In conclusion, the $P_4$ population observed on Titan could similarly be due to electron emission from the aerosols heated by the ion chemistry and the collection of the numerous electrons present on the dayside. In particular, a previous experimental work done by Chatain et al. (2020) showed that analogues to Titan's aerosols are chemically modified by exposure to a laboratory plasma mimicking Titan's ionosphere, which could participate to the heating of the aerosols.

# 7 Conclusions

The re-analysis of the Cassini Langmuir probe dataset in the ionosphere of Titan below 1200 km led to a better understanding of this ionized environment. Paper I, which details the re-analysis method, discuss the presence of several electron populations that are not expected by the ionospheric models that predict only one thermalized electron population. Between 2 and 4 electron populations are detected depending on the solar illumination and altitude. In this second paper we analyzed the complete Langmuir probe dataset below 1200 km with the objective to characterize the four populations and find correlations through statistics.

A first population at very low density is attributed to photoelectrons or secondary electrons emitted by the probe boom. The second population is detected in all conditions, with a density ~500 cm$^{-3}$ and temperature of $0.04 \pm 0.02$ eV ($460 \pm 230$ K) at 1000 km altitude. It is attributed to background thermalized electrons, possibly originally coming from suprathermal magnetospheric electrons.

The two remaining populations are very sensitive to solar zenith angle and extreme UV fluxes. Their density also increases with pressure, suggesting that their formation processes deal with the deposition of solar photon energy into the ionosphere. They are the dominant electron populations on dayside. The third population is observed at all altitudes, on dayside (up to 2700 cm$^{-3}$) and in lower density on nightside near the terminator. Its temperature decreases with altitude, down to $0.06$ eV $\pm 0.02$ eV ($700 \pm 230$ K) at 1000 km. It is suspected to be closely linked to the photo-ionization of the gas phase on dayside. On the opposite, the fourth population is rigorously observed only on dayside and below 1200 km altitude, and its temperature is linearly correlated to its density (up to $0.12$ eV / 1390 K and 2000 cm$^{-3}$). Its origin could be related to the presence of dust exposed to solar photons and active ion chemistry, through photo- or thermo-emission. Both of these populations are much hotter than the thermalized electrons predicted by models (see discussion in the introduction). The formation processes of these two populations are currently not (or partially) taken into account in the models and including these would be a very interesting future study. In particular, it could increase the average temperature of ionospheric electrons in the models, and consequently decrease the discrepancy with electron temperature measurements.

## Acknowledgments and Data

The Swedish National Space Board (SNSB) supports the RPWS/LP instrument on board Cassini. A.C. acknowledges ENS Paris-Saclay Doctoral Program. O.S. acknowledges funding by the Royal Society grant RP\EA\180014. N.J.T.E. was funded by the Swedish National Space Board under contract 135/13 and by the Swedish Research Council under contract 621-2013-4191. N.C. acknowledges the financial support of the European Research Council (ERC Starting Grant PRIMCHEM, Grant agreement no. 636829).

All Cassini RPWS data are archived in the Planetary Data System (PDS) Planetary Plasma Interaction (PPI) node at https://pds-ppi.igpp.ucla.edu on a pre-arranged schedule. The authors thank the TIMED/SEE and SORCE/SOLSTICE instruments' teams for providing the solar irradiance data, available at http://lasp.colorado.edu/lisird/data/timed_see_ssi_l3/ and http://lasp.colorado.edu/lisird/data/sorce_ssi_l3/, respectively.